\documentstyle[preprint,aps]{revtex}
\tightenlines
\begin{document}
\draft
\title
{\bf Exact $^4$He Spectral Function\\
in a Semirealistic NN Potential Model}
\author{Victor D. Efros$^{1)}$, Winfried Leidemann$^{2,3)}$, and
Giuseppina Orlandini$^{2,3)}$}
\address{
1) Russian Research Centre "Kurchatov Institute", Kurchatov Square 1,
123182 Moscow, Russia\\
2) Dipartimento di Fisica, Universit\`a di Trento, 
 I-38050 Povo (Trento), Italy\\
3) Istituto Nazionale di Fisica Nucleare, Gruppo collegato di Trento, Italy
}

\date{\today}
\maketitle
\begin{abstract}
The spectral function of $^4$He is calculated with the Lorentz integral
transform method in a large energy and momentum range. The excitation 
spectrum of the residual 3N--system is fully taken into account. The 
obtained spectral function is used to calculate the  quasi elastic 
longitudinal $(e,e')$ response $R_L$ of $^4$He for $q=$ 300, 400, and 500 
MeV/c. Comparison with the exact $R_L$ shows a rather sizeable disagreement 
except in the quasi elastic peak, where the differences reduce to about 10\% 
at q=500 MeV/c. It is shown as well that very simple momentum distribution 
approximations for the spectral function provide practically the same results 
for $R_L$ as the exact spectral function.
\end{abstract}
\bigskip
\pacs{PACS numbers: 
25.30.Fj, 21.45.+v, 21.10.Jx, 27.10.+h\\}

\vfill\eject

Data on electromagnetic processes on nuclei can be analyzed in a very simple 
way with the help of a spectral function (SF). The approximations involved in 
such an analysis are few and transparent. There exists an extensive literature 
dealing with evaluations and applications of the SF to 
exclusive, semiinclusive or inclusive reactions 
\cite{CS96}. However, only for 3--body nuclei exact calculations of
the SF have been performed \cite{CPS80}. A complete 
evaluation is very difficult for $A>3$ since
it requires the knowledge of the complete set of eigenstates for the 
(A-1)--subsystem. In fact only the (A-1) ground state is often known 
accurately, while excited states, especially those belonging to the
continuum, are much less under control, if not completely unknown. Already 
for $^4$He one finds only approximate evaluations of the SF \cite{MS91},
where the final state interaction in the residual 3N--system is neglected.
So the quality of the approximations which make use of the SF is often 
obscured by its poor knowledge.

Applying the method of the Lorentz integral transform \cite{ELO94} one can 
reduce the complexity of the calculation of the SF considerably.
In the present work we use this method to calculate the full SF 
of $^4$He with the semirealistic Trento (TN) potential model (central
force describing $^1S_0$ and $^3S_1$ phase shifts up to pion threshold). 
The result obtained is then used to evaluate the PWIA longitudinal $(e,e')$ 
response function $R_L$ at intermediate momenta. The resulting 
$R_L$'s are compared with the exact ones from Ref. \cite{ELO97} for the same 
NN potential. Such a comparison enables us to draw conclusions 
about the precision of the SF ansatz in inclusive $(e,e')$ scattering within 
a non relativistic framework. Since $^4$He is the lightest tightly bound 
nucleus, these results may be significant also for more 
complex nuclei.

The spectral function $S(k,E)$ represents the joint probability of finding a 
particle with momentum ${\bf k}$ and a residual (A-1)--system with energy $E$. 
The momentum {\bf k} and the energy $E$ are taken with respect to the c.m. 
and the ground state of the A--system, respectively: 
\begin{equation}
S(k,  E)={1\over 2 J_0+1}\sum_{f, s_z, t_z, M_0
}|\langle\psi_f^{A-1};{\bf k}s_zt_z|\psi_0^A(J_0M_0)\rangle|^2
\,\delta(E-(E_f^{A-1}-E_0^A))\,\,.
\end{equation}
Here $s_z$ and $t_z$ are the third components of the particle spin 
and isospin; $E_f^{A-1}$ and $\psi_f^{A-1}$ are eigenvalues and eigenstates
of the (A-1)--system; $J_0,M_0$ and $E_0^A$ are the total angular momentum, its 
third component and the ground state energy of the A--system, respectively.
There is a certain number of sum rules the SF has to fulfill:
\begin{equation}
\int d{\bf k}\, dE \,S({k}, E) = \int d{\bf k}\,n(k) = \,1\,,
\end{equation}
\begin{equation}
{1\over 2m}\int dE\, d{\bf k}\, {\bf k}^2\,S(k, E) = \langle T \rangle \,,
\,\,\,
\langle E\rangle= {A-2\over A-1}\,  \langle T \rangle-\,2\,{E_0^A\over A}   \,.
\end{equation}
Here $n(k)$ is the momentum distribution of the A--particle system
and $\langle T \rangle$ is the mean kinetic energy of a particle in the 
ground state. The last relation in Eq. (3) is the so called Koltun sum rule 
for the mean separation energy \cite{K74B71}. 
These sum rules form a set of constraints to test the accuracy of a 
calculation of $S(k, E)$.

In the following we will consider the proton spectral function $S_p(k,E)$. 
In this case the first two sum rules of Eqs. (2,3) have to be modified by 
an additional factor $Z/A$ on the right--hand sides.

In order to express the one--body knock out cross section in terms of the
SF two approximation are required:
(i) the particle interacting with the external probe is the one 
detected in experiment, (ii) this particle does not interact with the 
residual (A-1)--system (PWIA). With these two assumptions the exclusive or
semiinclusive one--body knock out cross sections can be written in 
a factorized form 
$ \sigma\simeq C A\sigma_N S(|{\bf k_f}-{\bf q}|,E)$.
Here $\sigma_N$ is the elementary cross section of the knocked out 
particle, ${\bf k_f}$ is its momentum in the lab--system,
${\bf q}$  is the momentum transfer, $E$ is the missing energy, and $C$ is a
kinematical factor. The so called {\it  quasi elastic} (q.e.) cross section 
can also be written in terms of $S(k,E)$ under the above assumptions. In 
particular the longitudinal response  entering the $(e,e')$ cross section 
reads  
\begin{equation}
R_L(q,\omega)\simeq A\tilde G_p^2(q_\mu^2) \int d{\bf k_f}\,dE\, S_p
(|{\bf k_f}- {\bf q}|,E)\, \delta\left(\omega- E-{{\bf k_f}^2\over 2m} -
{{\bf k}_{A-1}^2\over 2M_{A-1}}\right) \,,
\label{R_L}
\end{equation}
where $\omega$ is the energy transfer, $q^2_\mu = q^2 - \omega^2$, $\tilde G_p$
is the free proton electric form factor \cite{DF84}, while $M_{A-1}$ and 
${\bf k}_{A-1}={\bf q}-{\bf k_f}$ are mass and recoil momentum of the 
(A-1)--system, respectively. Here we do not consider an off-shell nucleon form 
factor, since our aim is a consistent comparison to the full $R_L$ of Ref. 
\cite{ELO97}, where such effects were not considered. 
The definition above includes only the proton responses of the nucleus.
In principle one has also to consider the neutron responses, but at low and 
intermediate $q$ the neutron electric form factor is negligible
($(\tilde G_n/\tilde G_p)^2 \simeq 1$ \% at $q_{\mu}^2= (500$ MeV/c$)^2$). 

The SF can be calculated with the Lorentz integral transform method 
\cite{ELO94} as already pointed out in Ref. \cite{ELO95}. Let us first denote 
the overlap of the $A$--body bound state with the single nucleon plane wave
\begin{equation}
\chi_{p/n}({\bf k};s_z, M_0) = \langle {\bf k},s_z, t_z=\pm 1/2
|\psi_0^A(J_0M_0)\rangle\,.
\label{rhs}
\end{equation} 
It represents a localized state in the subspace pertaining to the residual 
(A-1)--subsystem. Written in terms of this quantity,
the proton SF
\begin{equation}
S_p(k,E)={1\over 2 J_0+1}\sum_{f, s_z, M_0}|\langle\psi_f^{A-1}|
\chi_p({\bf k};s_z,M_0)\rangle|^2 \delta(E-(E_f^{A-1}-E_0^A))\,, 
\end{equation}
looks similar to a response function of the (A-1)--subsystem with 
$\hat{O}\psi_0^{A-1}$ replaced by $\chi_p$. Therefore we can proceed by 
analogy with the calculation of a response function. We obtain $S_P(k,E)$
as a solution to the integral equation 
\begin{equation}
\int {S_p(k,E) \over (E - \sigma_R)^2 + \sigma_I^2}\,dE=\Phi_p
(k,\sigma_R,\sigma_I) 
\label{tra}
\end{equation}
whose right--hand side is given by
\begin{equation}
\Phi_p(k,\sigma_R,\sigma_I)={1\over 2 J_0+1}\sum_{s_z, M_0}
\langle\tilde\Psi_p({\bf k};s_z,M_0)|
\tilde\Psi_p({\bf k};s_z,M_0)\rangle\,,
\end{equation}
where $\tilde{\Psi}_p$ is a localized solution to the 3--body inhomogeneous 
equation
\begin{equation}
(H_{A-1}-E_0^A-\sigma_R+i\sigma_I)\tilde\Psi_p({\bf k};s_z,M_0)=\chi_p
({\bf k}; s_z,M_0)\,. 
\label{diff}
\end{equation}
We solve Eq. (\ref{diff}) expanding $\tilde\Psi_p$ in hyperspherical harmonics. 
Complete convergence of the expansion is reached with similar values for the 
expansion parameters as in Ref. \cite{ELO97}. The inversion of the Lorentz 
integral transform, Eq. (\ref{tra}), is carried out as described in Ref. 
\cite{ELO94}. Quite a good stability of the inversion results is observed. 
As previously we check the quality 
of the results with the help of sum rules as well. To this end we evaluate 
the sum rules of Eqs. (2,3) by an explicit integration of the properly 
weighted calculated SF. We obtain the following relative differences 0.9 \% 
(norm), 0.2 \% ($\langle T \rangle$), and 0.8 \% (Koltun sum rule assuming 
that $S_n(k,E) = S_p(k,E)$). Since the sum rules weight $S(k,E)$ in  
different regions these results point out that the SF is calculated with a  
satisfying precision.

Before coming to the SF, in Fig. 1 we show the $n(k)$ of $^4$He for 
the TN potential in comparison to that obtained for a realistic potential 
(Argonne $v_{18}$ + Urbana IX)  \cite{Wpc}. One sees a rather good agreement 
up to almost 2 fm$^{-1}$. However, different from the realistic result 
the semirealistic $n(k)$ is considerably smaller at higher k. Most of
these differences are presumably explained by the missing tensor force
in the TN potential (see Ref. \cite{MAT88}). We also show in 
Fig. 1 two partial momentum distributions. They are obtained from Eqs. (1,2)
if the sum over $f$ in Eq. (1) is restricted either to the triton ground 
($n_{tp}(k))$ or its continuum state ($n_{t^*p}(k))$. As expected (see Ref. 
\cite{CS96}) $n_{tp}(k)$ governs the lower $k$ momentum distribution, 
while $n_{t^*p}(k)$ dominates at higher $k$. The integration of 
$n_{tp}(k)$ leads to the so called spectral factor. For the TN potential one 
finds a spectral factor of 0.89, whereas with the above realistic potential a 
value of 0.84 \cite{Wpc} is obtained. 

In Fig. 2 we show $S_p(k,E)$. Only energies above the breakup threshold 
$E^{A-1}_{thr}$ of the rest nucleus are illustrated, while
the contribution from the bound state of the rest nucleus is identical 
to the $n_{tp}(k)$ of Fig. 1 (see Eq. (1)). The values of
 $S(k,E\ge E^{A-1}_{thr}+1$ MeV) are plotted in the figure. We note that 
$S(k,E^{A-1}_{thr}) = 0$, and thus $S(k,E)$ exhibits a rather strong 
slope at low energy. 
For momenta below 2 fm$^{-1}$ one finds a sharp maximum at about 2 MeV 
above $E_{thr}^{A-1}$. On the contrary $S(k,E)$ is flat in most 
other regions. Only for $k > 2$ fm $^{-1}$ there is a ridge where the 
peak position shifts to higher $E$ for increasing $k$.

As already mentioned one of our aims is a comparison of the  q.e. $R_L$
with the exact one in an intermediate  $q$ range for the same NN potential.
In Fig. 3 we show both $R_L$'s in comparison to experimental data. We would 
like to point out that the full results are a bit different from those in Ref.
\cite{ELO97} for two reasons:
(i) in the calculations of Ref. \cite{ELO97} $G_n(q_{\mu}^2)$ entered 
erroneously with a negative sign leading to small -- but not totally 
negligible -- effects on $R_L$ (e.g., peak height and high--energy 
tail become a bit lower); 
(ii) different from Ref. \cite{ELO97} here we account for the small
overbinding of the TN potential for $^4$He. The threshold energy reads 
$\omega_{thr} = E_0(^3{\rm H}) - E_0(^4{\rm He}) + q^2/2M(^4{\rm He})$
and we correct the overbinding 
by shifting our response to lower energies to make $\omega_{thr}$ correspond 
to the one with the experimental $E_0(^4$He).

Of course these two modifications do not change the general picture given in 
Ref. \cite{ELO97}. The good agreement with experiment becomes even better for 
q.e. peak and high-energy tail. At low energy there is a slight 
improvement at $q = 300$ MeV/c.
For the two higher $q$ the agreement with experiment in the threshold region
is still satisfying but not as excellent as shown in Ref. \cite{ELO97}. 

Figure 3 shows, as expected, that at $q = 300$ MeV/c the PWIA does not lead 
to a good description of $R_L$. The q.e. peak is shifted to higher energies 
by 15 MeV and the peak height is overestimated by more than 40 \%. The 
overestimation becomes worse with increasing energy, while at low energy
$R_L$ is underestimated. At the two higher $q$ the peak is shifted by 12 MeV, 
but since the peak width grows with increasing $q$ this shift is a minor 
effect. The shift of the peak can be qualitatively understood considering a 
nucleon at rest in a potential well: $\omega$ can be estimated as 
$q^2/(2m)+ V_f - V_i$, where $V_{i,f}$ are the potential energies before or 
after interaction with the virtual photon. While $V_f$ is negative, it becomes 
zero in  PWIA leading to an increase in $\omega$. The peak height improves with 
overestimations of 25 \% at $q = 400$ and 13 \% at 500 MeV/c. Thus one has 
to expect that beyond 500 MeV/c the PWIA is a good approximation at 
the q.e. peak. Beyond the peak the PWIA result still overestimates the exact 
one considerably, but the discrepancy decreases with increasing $q$. At low 
energy, however, the underestimation remains considerably large.

It is advantegeous to have a simple and good approximation for the PWIA 
response. From Fig. 2 it is evident that at low $k$ almost all the strength of 
$S(k,E)$ with the disintegrated rest nucleus is found close to the breakup 
threshold. This suggests the following approximation 
\begin{equation}
S(k,E) \simeq n_{tp}(k)\delta(E-E_0(^3{\rm H}) + E_0(^4{\rm He}))
+ n_{t^*p}(k)\delta(E-E_{t^*p} + E_0(^4{\rm He})) 
\end{equation}
for calculating inclusive processes, where $E_{t^*p}$ is the breakup energy of 
the rest nucleus. One obtains an even simpler approximation considering 
that $E_{t^*p} \simeq E_0(^3{\rm H})$:
\begin{equation}
S(k,E) \simeq n(k) \delta(E-E_0(^3{\rm H}) + E_0(^4{\rm He})) \,.
\end{equation}
Eq. (11) was used e.g. in Ref. \cite{CS92} where large deviations from
the full GFMC response for a realistic potential at $q=400$ MeV/c were
reported. In Fig. 4 we show the PWIA results with the above two approximations 
at $q=500$ MeV/c relative to the full SF result. It is readily seen that the 
three responses are very similar, particularly in the q.e. peak region 
(at $q=300$, 400, and 1000 MeV/c one has very similar results). It is worth 
mentioning that our PWIA result at $q=400$ MeV/c is essentially the same as 
the one in Ref. \cite{CS92}. This shows again that a semirealistic central 
force leads for $R_L$ practically to the same result as a realistic potential.
 
In this work we obtain for the first time the full spectral function of 
$^4$He. A semirealistic NN potential is used. The final state interaction in 
the residual system is taken into account completely by the Lorentz integral 
transform method. The SF is then used to calculate the  q.e. longitudinal 
response function of $^4$He which is compared to the exact one of Ref. 
\cite{ELO97}. In the peak 
region the differences decrease with growing momentum transfer up to 
about 10 \% at $q=500$ MeV/c, but one still finds sizeable differences apart 
from the peak. Similar results were found in Ref. \cite{Gl94} for $^3$H and
$^3$He for the studied momentum transfers of 300 and 400 MeV/c. Different from 
the 3--body system $^4$He  already resembles some aspects of more complex 
nuclei and thus the general picture of the q.e. response should not change much
in such systems. We show as well that the simple momentum distribution 
approximations for the SF provide results for $R_L$ which are quite close to 
those obtained with the full SF.

Two of us (W.L. and G.O.) thank the Institute of Nuclear Theory at 
the University of Washington for its hospitality and the Department of Energy 
for partial support during the completion of this work. The work of V.D.E. was
supported by INFN and RFBR (grants no 96-15-96548  and 97-02-17003).

\begin{figure}
\caption{ 
Total (solid curve) and partial momentum distributions $n_{tp}$ (dotted curve) 
and $n_{t^*p}$ (dashed curve) of $^4$He with TN potential; also shown total 
result (full dots) and $n_{tp}$ (open squares) with Argonne $v_{18}$ + Urbana 
IX [9].}
\end{figure} 

\begin{figure}
\caption{$S_p(k,E)$ of $^4$He with TN potential in units of
fm$^{3}$MeV$^{-1}$.}
\end{figure} 

\begin{figure}
\caption{$R_L$ of $^4$He with TN potential:
PWIA results according to Eq. (4) (dashed curves) and 
full results (solid curves); experimental data from Bates [11] and Saclay 
[12].} 
\end{figure} 

\begin{figure}
\caption{$R_L$ with SF of Eq. (10) (dashed curve) and of Eq. (11)
(dotted curve) relative to $R_L$ with full SF (the q.e. peak is marked
by an arrow).} 
\end{figure}
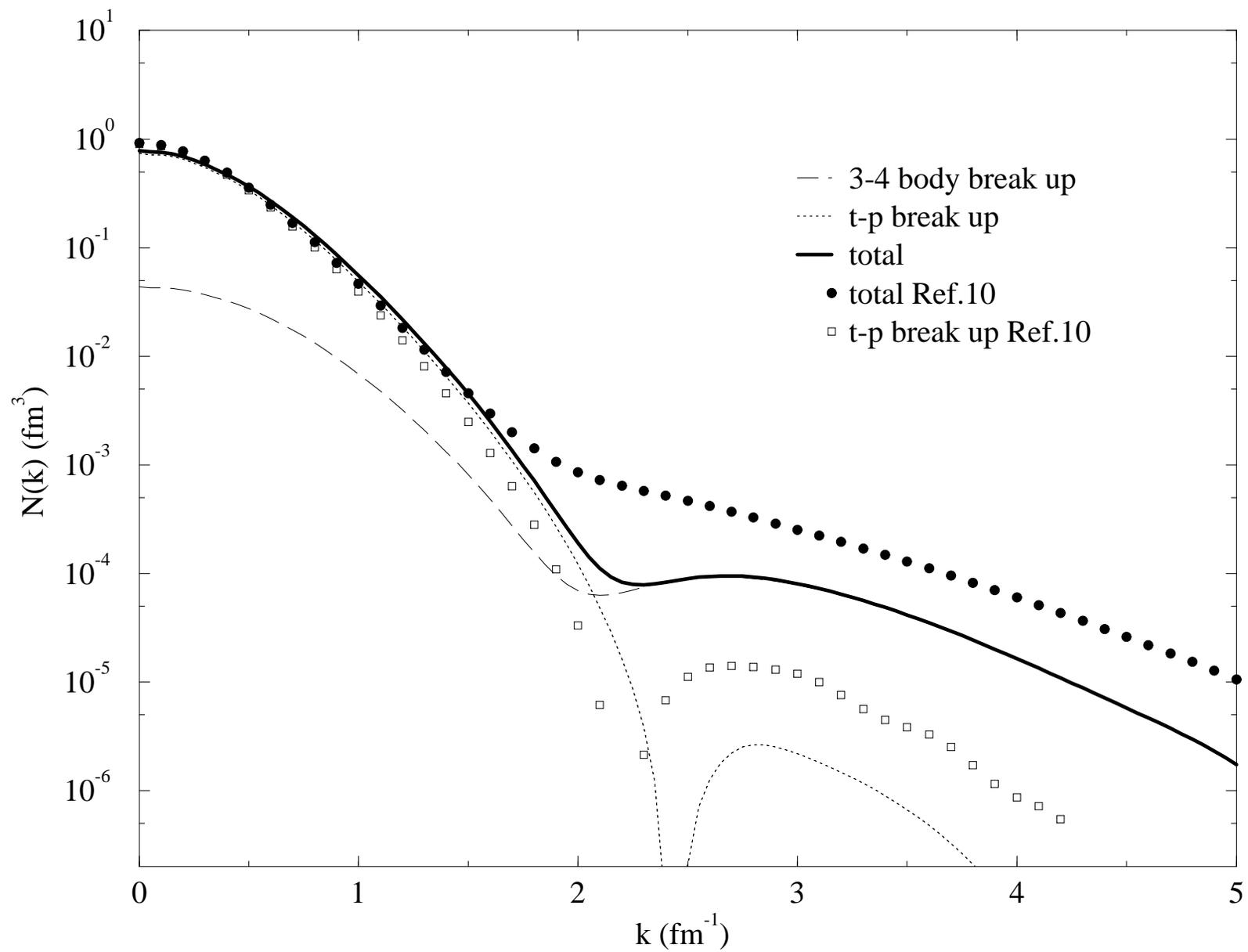
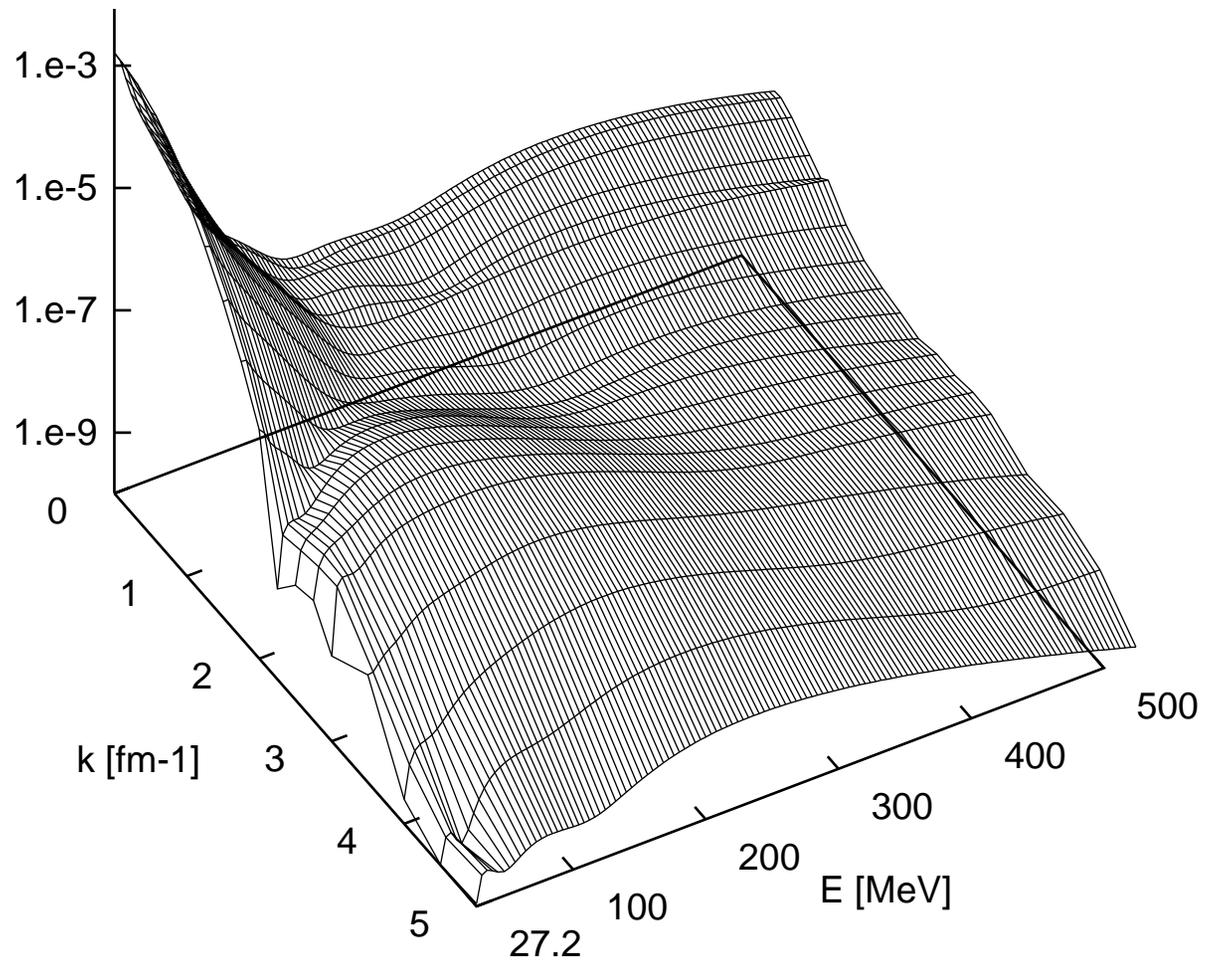
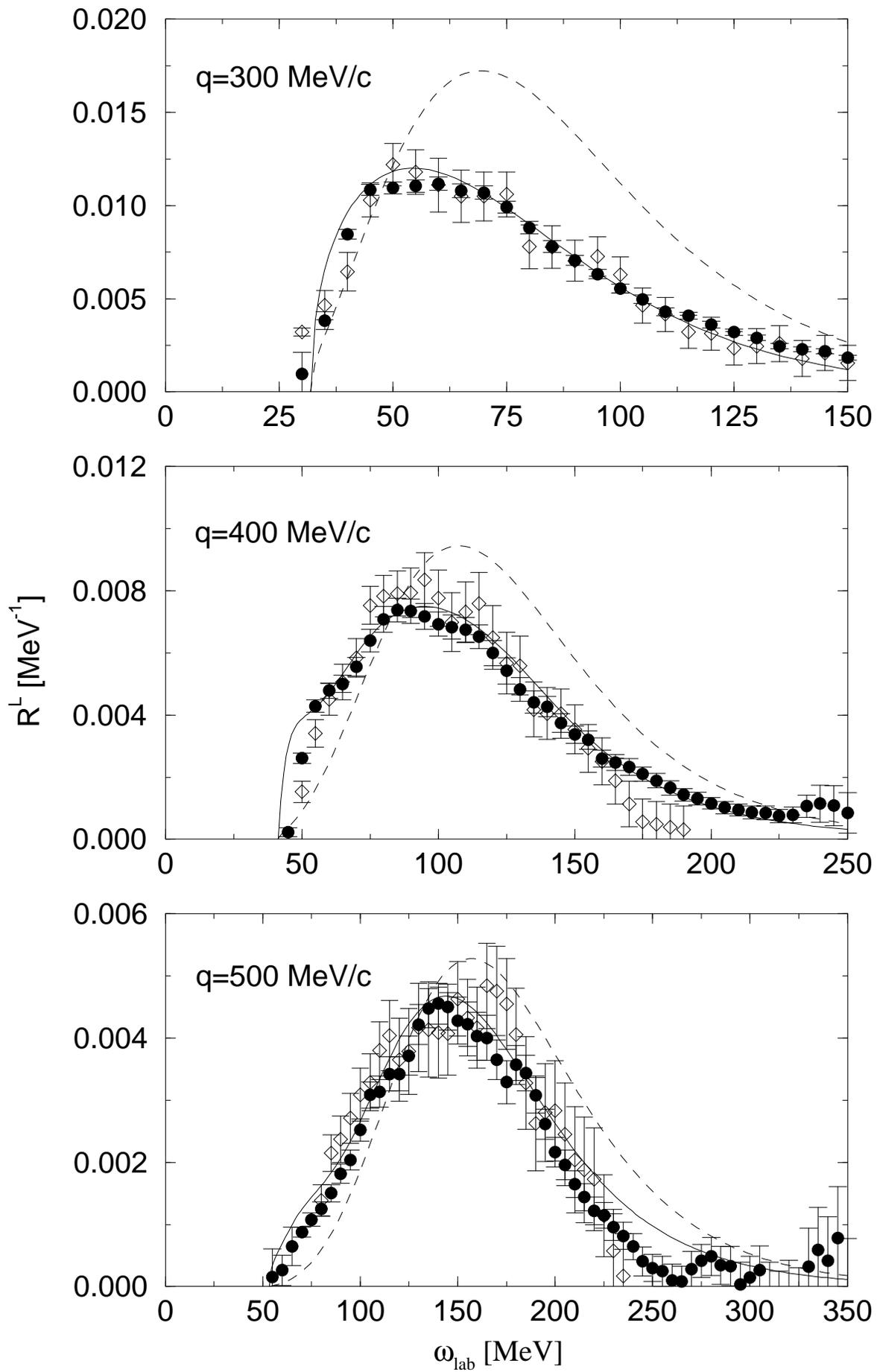
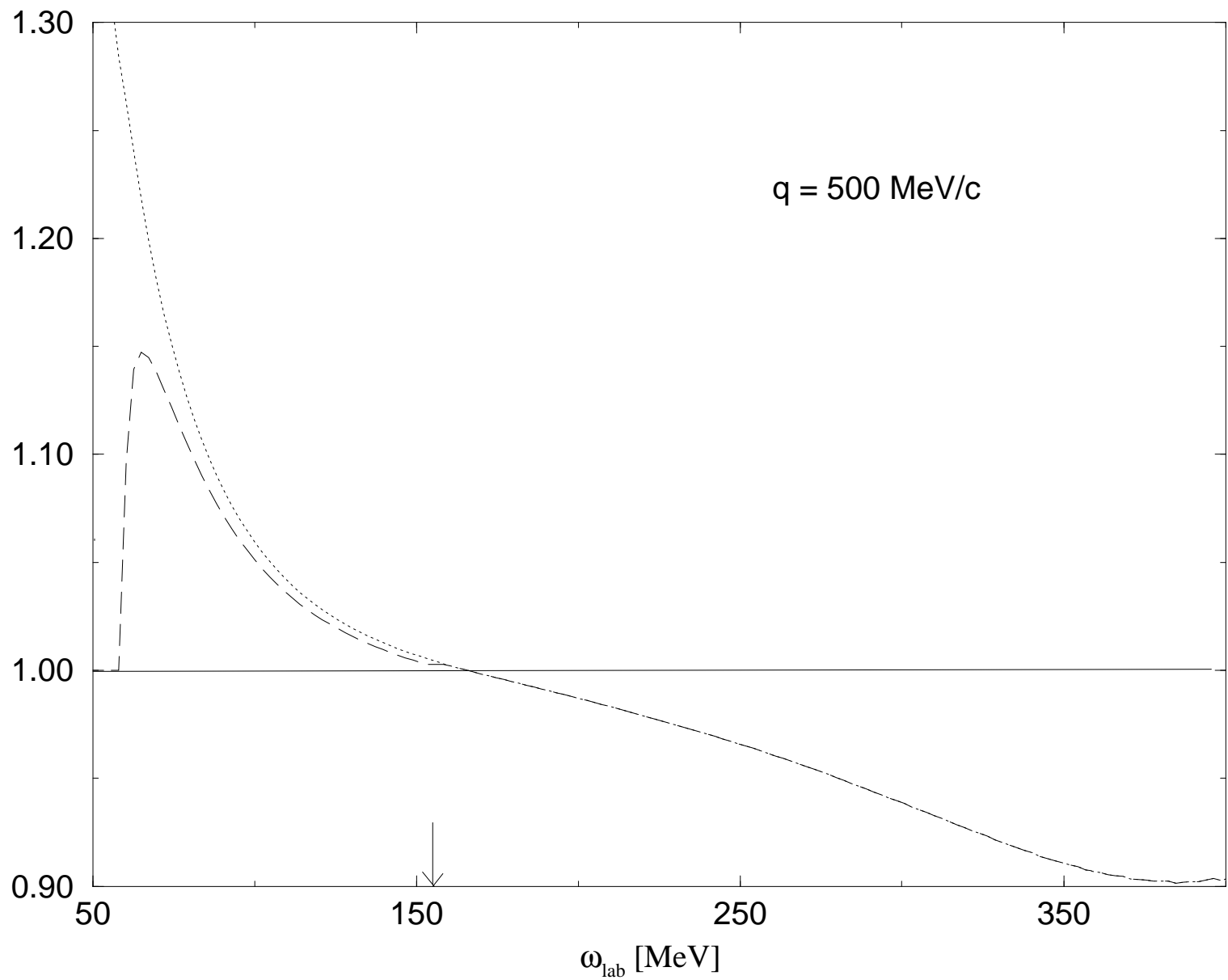
\end{document}